# The effect of fluorine or chlorine substitution on mesomorphic properties of ferroelectric nematic liquid crystals

Martin Cigl,[1] Natalia Podoliak,[1] Dalibor Repček,[1] Pavlo Golub,[1] Marta Lavrič,[2] and Vladimíra Novotná [1]*

[1] *Institute of Physics of the Czech Academy of Sciences, Na Slovance 1999/2, 182 00 Prague 8, Czech Republic*

[2] *Jožef Stefan Institute, Jamova cesta 39, SI-1000 Ljubljana, Slovenia*

Corresponding author*: Vladimíra Novotná, e-mail: novotna@fzu.cz

Key words: liquid crystals, ferroelectric nematics, SHG measurements, dielectric spectroscopy

## Abstract

Ferroelectric nematic phase ($N_F$) represents an attractive and foremost field of liquid crystals, combining fluidity with ferroelectricity. $N_F$ materials exhibit large polarization values and remarkable non-linear optical properties. We have designed an original molecular structure with halogen substituents in the position of an electron donating group. In a prolonged molecular core, such a modification led to the presence of the ferroelectric nematic phase ($N_F$) below the nematic one. Besides, an application of Cl atom in the molecular core of one of the presented materials has been utilized for the first time for ferroelectric nematogens. We have examined mesogenic behaviour and ferroelectric characteristics of the $N_F$ phase. In the $N_F$ phase for the cell with antiparallel rubbing, we have detected a textural transformation, which evidences strong polar character of anchoring at the surfaces. The presented results provide valuable insight into the design of ferroelectric nematic liquid crystalline compounds.

## 1    Introduction

Since the discovery of a ferroelectric nematic phase ($N_F$), an intensive development has started in the field of liquid crystals. Even though such a phase was predicted more than a hundred years ago,[1] the first experimental examples were found in 2017, when two compounds designated DIO and RM734 were described to exhibit ferroelectric properties in a nematic phase.[2-4] In the ferroelectric nematic phase, the dipole moments must be sufficiently strong to overwhelm thermal fluctuations.[5-8] Contrary to conventional nematic phase (N), in which the director orientations are indistinguishable, and the long-range order has an axial character, in the $N_F$ phase the molecular directors are arranged in a polar manner.

Recently, several reviews were published to summarize up-to-date knowledge and outline prospects for further research.[9-11] Progress in studies of ferroelectric nematic phase depends on the design of new compounds and on the detailed analysis of the observed phenomena. $N_F$ phase represents a polar fluid with huge electro-optical response, dielectric constant, piezoelectric coupling and non-linear optical coefficients.[12,13] However, the dielectric spectroscopy is still a subject of an intensive debate.[14-16] as the measurements of permittivity include an interfacial of nematogen and surfactant layer. The recent analysis of dielectric measurements supports the idea that the permittivity in the $N_F$ phase is indeed huge, and the presented model can explain the unusual temperature dependence close to the N-$N_F$ phase transition.[17]

Depending on the anchoring conditions, characteristic textures with twisted domains are observed in the ferroelectric nematic phase.[18-21] Recently, topological aspects of the $N_F$ phase have been pursued, and a variety of polar topological structures has been observed, as well as transformations between diverse configurations have been described in dependence on a cell-surface alignment.[22-24] For a range of ferroelectric nematogens, an additional phase has been found between the regular nematic phase (N) and the $N_F$ phase, revealing an antiferroelectric character.[25-27] This intermediate phase, labelled as $N_x$, $N_s$ or $SmZ_A$ phase [25,26] is still a subject of intensive discussions. It exhibits low values of permittivity, and polarization current shows a characteristic double peak. In polarizing microscope, characteristic zig-zag defects were observed, which supported lamellar character of such a phase.[27] This phase possesses only a short-range order and, contrary to traditional smectics, antiferroelectric domains are regularly distributed with boarders perpendicular to polar direction. Such a layered organization has been proven by a resonant x-ray.[26] A competing balancing mechanism due to the flexoelectric coupling between the electric polarization and splay deformation was proposed.[28] The presence of an intermediate phase was confirmed by precise calorimetry measurements.[29,30]

Synthetic effort is concentrated on the tailoring of the mesogenic properties, to follow requirements of stability and temperature range extension towards the room temperature.[31-38] The general molecular structures of ferroelectric nematogens should have suitable aspect ratio and large dipole moment, which is present due to the effective combination of an electron donating group (EDG) and an electron withdrawing group (EWG). Their importance is not only in creation of a sufficiently large dipole moment, but the influence of other molecular parameters on the molecular packing via various intermolecular forces should be taken into consideration. From the discovery of ferroelectricity in nematic phase, polar smectic phases [39-41] and chiral ferroelectric nematics [42-43] have also been described.

Fluorination is frequently used in the design of ferroelectric nematogens, and fluorine is logically applied in the positions, where it contributes to the total longitudinal electric dipole moment of the molecule.[44-45] To date, a nitro group seems to be an ideal moiety to serve as EWG in the design of new polar molecules. Recently, we investigated the effect of the exchange of oxygen-based EDG, used in most ferroelectric nematogens, for even more efficient nitrogen.[32] In the present research, we examine halogen substituents in the place of EDG. Such design seems counterintuitive as halogens are generally electronegative elements.

Nevertheless, halogens also represent highly localized electron density in the molecule, which might support the molecular organization into the polar $N_F$ phase. Besides, from the chemical reactivity of halogenated aromatic compounds, the halogen atom is known to partially share electron density from their lone electron pairs in such a way that it can to some extent compensate for the electron-withdrawing effect. Additionally, halogens are the only atoms abundant with lone electron pairs, which can be used without any further fragment like alkyl for oxygen, nitrogen or sulphur. Due to these properties, we aimed to investigate the effects of this substitution on the molecular self-organising behaviour and its surface anchoring.

## 2    Experimental

## 2.1.    Synthesis

Synthesis of fluoro-terminated materials started from commercially available 4-fluorosalicylic acid (**1**), which was converted to ester and alkylated with *n*-bromohexane (Scheme 1). After the basic hydrolysis of the methylester group of **2**, the acid **3a** was obtained. The chlorinated analogue was synthesized in two steps from 4-chloro-2-hydroxyacetophenone (**4**). The step was the introduction of the sidechain via alkylation reaction and the second comprised haloform reaction to transform the acetyl group to carboxylic giving acid **3b**. In the next step, the benzoyl chloride **6**, prepared as described previously,[32] was used in the esterification reaction with benzyl 4-hydroxybenzoate (**7**) and the protective methoxycarbonyl group was subsequently removed by means of aqueous ammonia in THF to yield benzyl-protected intermediate **8**. The halogenated acids **3a-b** were combined with the hydroxybenzoate **7** or **8** in a DCC-mediated esterification reaction, after which the benzyl protective group was removed by the hydrogenolysis on palladium. The obtained intermediates **9a-c** were esterified with 4-nitrophenol (**10**) using EDC coupling to give target mesogens F6, FF6 and ClF6, for the chemical formula see Figure 1. NMR data and other details of the synthetic procedures are in Supplemental file (SI).

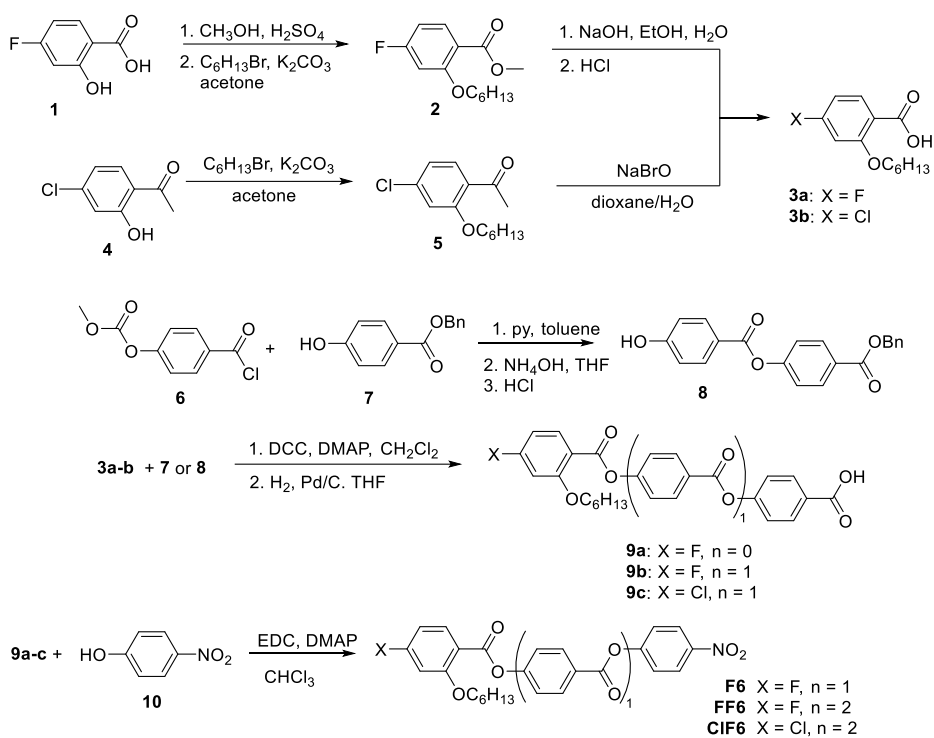

Scheme 1.    Synthetic route to target halogen-terminated mesogens.

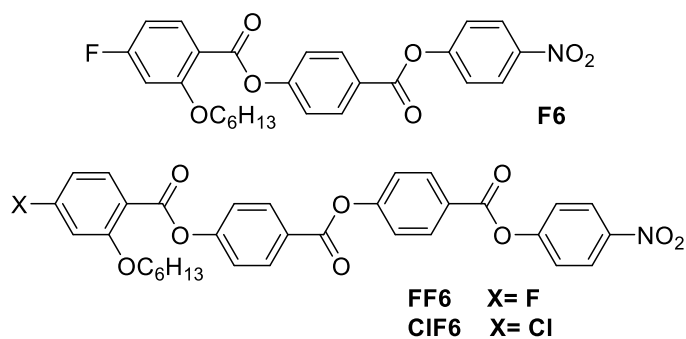

Figure 1.    The chemical formula of the target compounds F6, FF6 and ClF6.

## 2.2.    Measuring methods and set-up

For textural observations, commercial cells (WAT PPW company, Poland) with homogeneous geometry (HG) were utilized, with the rubbing direction of the surfactant layer parallel (HG-P) or antiparallel (HG-A) at the opposite surfaces. For comparison, the cell with homeotropic anchoring (HT) was used, in a geometry with molecules perpendicular to the cell surface. For the dielectric spectroscopy and polarization measurements, home-made 5-μm and 25- μm sandwich-type glass cells without surfactant were prepared. All the cells were filled with studied materials in the isotropic phase by means of capillary action. Polarized light microscopy (POM) observations were conducted using Nikon Eclipse E600 microscope, equipped with a heating-cooling stage Linkam with temperature stabilization with an accuracy of ±0.1 K.

Differential scanning calorimetry (DSC) measurements were performed using Perkin Elmer calorimeter. The samples of 2-5mg were hermetically sealed into aluminium pans and placed into the calorimeter chamber infiltrated with nitrogen. The measurements were performed during heating and cooling cycles at a rate of 10 K/min. For high-resolution AC calorimetric measurements, a home-made experimental setup has been utilized. A detailed description of the technique was given previously.[46] The thermal stability was better than 0.1 mK and the apparatus allowed us to perform very slow cooling and heating runs.

Dielectric spectroscopy studies were conducted with an impedance analyser. The sample temperature was stabilised within ±0.1 K during the frequency sweeps (1 Hz ÷ 1 MHz). From the resistance and capacity of the sample, real and imaginary parts of the complex permittivity, $\varepsilon*(f) = \varepsilon' - i\varepsilon''$, were obtained and the results were fitted to the Cole-Cole formula. Polarization, $P$, measurements were performed under the triangle-profile electric field with the frequency of 10 Hz and the magnitude of 10 V/μm, supplied by a generator. A switching current at the stabilized temperature was detected using a digital oscilloscope and was integrated to supply $P(T)$ values.

The second harmonic generation (SHG) measurements were performed in HG-P glass sandwich cells, with the cell thickness of 9-10 μm. SHG signal was detected in a transmission configuration with an optical set-up based on femtosecond Ti:sapphire laser (Spitfire ACE), with the central wavelength of 800 nm. The rubbing direction was oriented parallel to the polarization direction of the incident beam. Birefringence measurements were performed on thin 1.6-μm HG-P cells in a set-up based on a photoelastic modulator and equipped with a green light (λ = 550 nm) source. The birefringence, Δn, was calculated from the retardation. Details for all experimental techniques are in Supplemental file.

3. **Results and their discussion**

The target molecules (Figure 1) were prepared and characterized according to the procedures described above and in Supplemental file (SI) in details. We studied and compared two kinds of molecular structures, with three benzene rings (F6 compound) and four benzene rings (FF6 and ClF6) in the molecular core. Compounds FF6 and ClF6 differ only in the halogen at the terminal position of the molecule, FF6 is ended by a fluorine atom and ClF6 by a chlorine atom. The transition temperatures and phase identification were established based on the differential scanning calorimetry (DSC) in combination with the observations under a polarized light microscope (POM) and confirmed by other experiments. The DSC data are presented in Table 1 and the thermographs are shown in Figure S1 and Figure S2 in SI for FF6 and ClF6, respectively. Three-ring compound F6 is non-mesogenic and crystallization was not detected during the following cooling-heating runs. On contrary, four-ring FF6 and ClF6 compounds show a nematic (N) - ferroelectric nematic ($N_F$) phase sequence on cooling from the isotropic phase (Iso).

Table 1. The melting points, m.p. were taken on the second heating run, the phase transition temperatures and the crystallization temperature, $T_{cr}$, were taken subsequently on the second cooling from the isotropic phase (Iso). The temperature $T_{iso}$ corresponds to the Iso-N and $T_c$ to the N-$N_F$ phase transition. All temperatures are in °C, and the corresponding enthalpy changes, ΔH, in kJ/g, are in square brackets. Symbol * means that compound F6 does not show any crystallization during DSC runs.

|        | m.p./°C [ΔH/ Jg$^{-1}$] | $T_{cr}$./°C [ΔH/ Jg$^{-1}$] |       | $T_c$./°C [ΔH/ Jg$^{-1}$] |   | $T_{iso}$./°C [ΔH/ Jg$^{-1}$] |
|--------|-------------------------|------------------------------|-------|---------------------------|---|-------------------------------|
| F-6    | 76.8 [+85.8]            | *                            | -     |                           | - |                               |
| FF-6   | 124.9 [+65.0]           | 46.9 [-30.2]                 | $N_F$ | 102.5 [-0.20]             | N | 178.9 [0.46]                  |
| ClF-6  | 143.6 [+76.8]           | 52.2 [-27.8]                 | $N_F$ | 130.4 [-0.38]             | N | 201.2 [0.70]                  |

The phase behaviour of the materials was investigated by POM observations in various cells and confinement conditions. On cooling from the Iso phase, we identified textural changes and the phase transition temperatures. In both parallel and antiparallel HG cells, the nematic phase showed regular textural features, with highly birefringent perfectly aligned area with an extinction along the rubbing direction (Figure 2(a)). A schlieren texture was observed in the HT cell (Figure 3(a)). Nevertheless, for a cell without surfactant, the behaviour in the nematic phase can be different on electrode surfaces. It is demonstrated for a 5-μm home-made cell without surfactant for compound ClF6 in Figure 4 and for FF6 in Figure S3 (SI). For this type of the cell, in the N phase, the textures are completely black under the electrode area, showing no extinction when rotating the sample (Figure 4(a)).

In the HT geometry during cooling from the nematic phase, the schlieren texture was transformed (Figure 3(b)) and later continuously changed its character, see Figure 3(c). In HG geometry for both types of the cell alignment, strong pretransition effects are observed at the N-$N_F$ phase transition, and tiny coloured stripes appeared along the rubbing direction with enhanced director fluctuations (Figure 2(b)). In the HG cell with antiparallel rubbing, there is a temperature interval below the N-$N_F$ phase transition, where aligned texture with an extinction position along rubbing direction persisted, as is demonstrated in Figure 2(c). Such a regular texture with an extinction is unstable, and twisted domains start to grow. Such transformation has a character of a boundary, and a borderline between the unstable aligned texture and a twisted-domain state is seen in Figure 2(d). The video taken on cooling from the nematic phase in a 5-μm HG-A cell is in Supplemental file (SI).

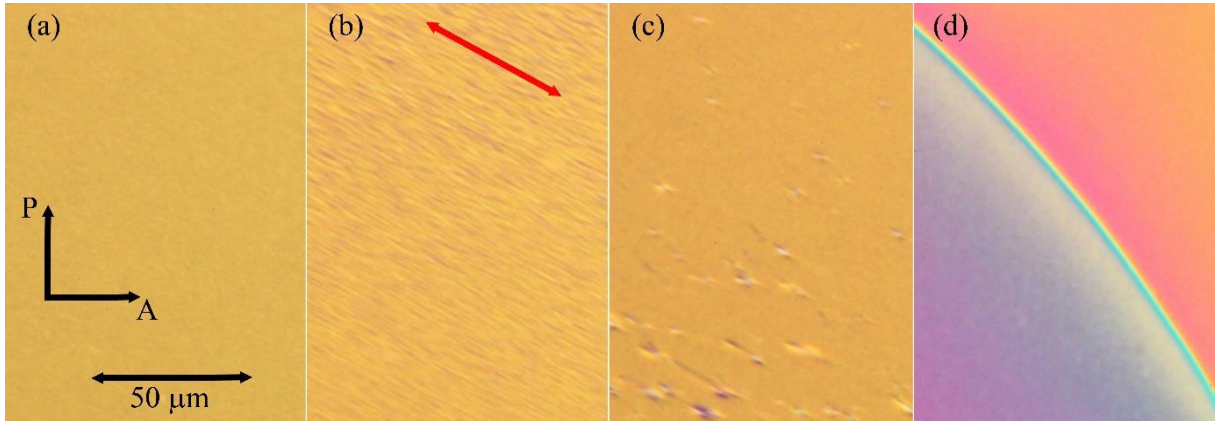

Figure 2. Compound FF6 in a 5-μm HG-A cell (with antiparallel rubbing in the direction marked by the red arrow) (a) in the N phase at T=105°C, (b) tiny stripy texture at the N-$N_F$ phase transition at T=102.5°C, (c) in the $N_F$ phase at T=102°C and (d) at T=101°C, with a front of texture transformation to twisted state (blue part coming from the left bottom corner).

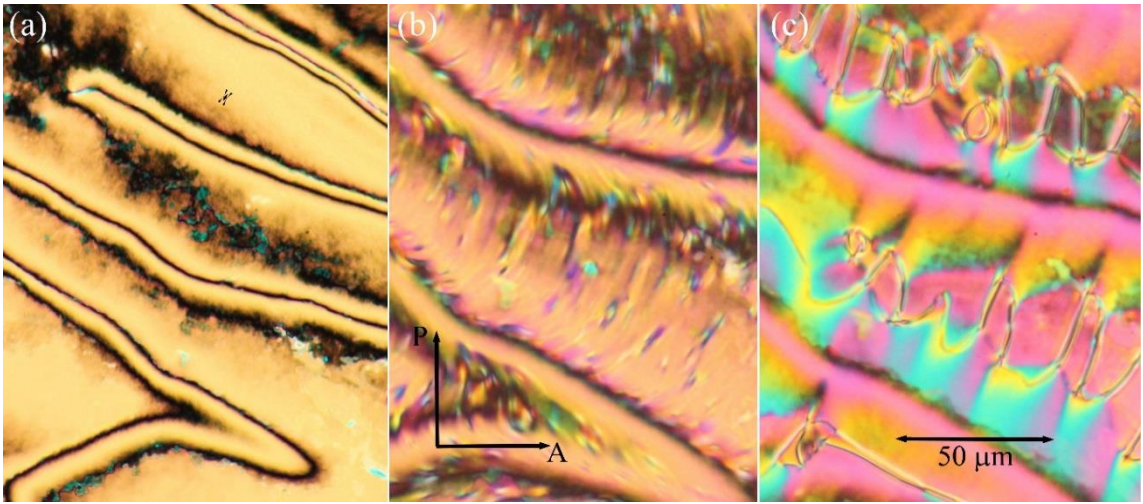

Figure 3. Textures for the compound FF6 in a 5-μm HT cell in (a) the nematic, (b) at the N-$N_F$ phase transition and (c) in the $N_F$ phase.

In the HG cell with antiparallel rubbing, twisted domains are usually observed in the $N_F$ phase being stable till crystallization. For the studied compounds, these domains prefer to be only slightly inclined from perpendicular direction with respect to the rubbing direction, see Figure 5 for ClF6 and Figure S3 for FF6. They were named "sierra-shaped" in Ref. 21 due to their mountains profile. The domains with opposite orientation of the twist can be distinguished when decrossing the polarizer position at an angle of about 20 degrees (Figure 5(b)) and Figure 5(c)) from the crossed position (Figure 5(a)). Nevertheless, we observed also another type of domains, oriented along the rubbing direction, often having a lenticular shape. In Figure 5, such a domain crosses the disclination line separating dominating type of twisted domains. The textures in HG-P cell show stripes oriented parallelly to the rubbing direction, for illustration see Figure S4 in SI. Such a kind of domains parallel to the rubbing direction can be occasionally found also in HG-A cell (Figure 6). In Figure 6 we demonstrate the situation in HG-A cell at its outer area, where the sierra-type twisted domains are seen in coexistence with a pseudo-regular array of stripes. Similar type of regular domains has been described in the $N_F$ phase for another materials, nevertheless, in a different geometry or conditions.[47] Detailed analysis of these domains and their growth is beyond the scope of this paper and will be published elsewhere.

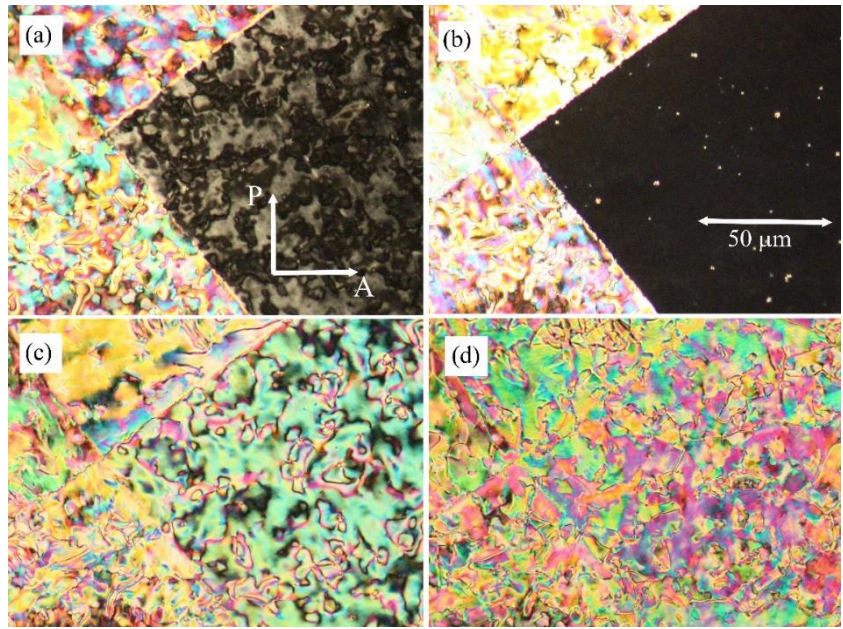

Figure 4. Textures for compound ClF6 in a home-made cell without surfactant, on cooling from (a) the nematic phase, T=140°C, (b) at the N-$N_F$ phase transition T=131°C, (c) in the $N_F$ phase at T=129°C and (d) T=125°C. Right part of all Figures is the area between electrodes, which looks dark in (a) and (b) pictures.

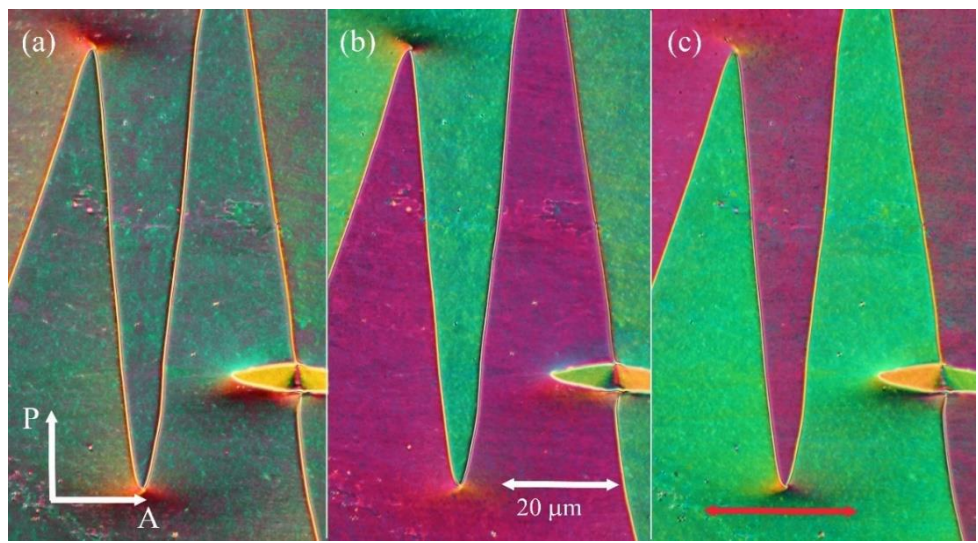

Figure 5. Twisted domains for ClF6 in the $N_F$ phase at T=40°C, in a 5-μm HG-A cell with antiparallel rubbing direction, marked by the red arrow in Figure (c). Textures taken (a) with the crossed position of polarizers; (b) and (d) show the same view with uncrossed position, when the polarizer (P) is rotated at an angle of about 20 degrees anticlockwise or clockwise from the analyser (A). Boarders between domains correspond to disclination lines.

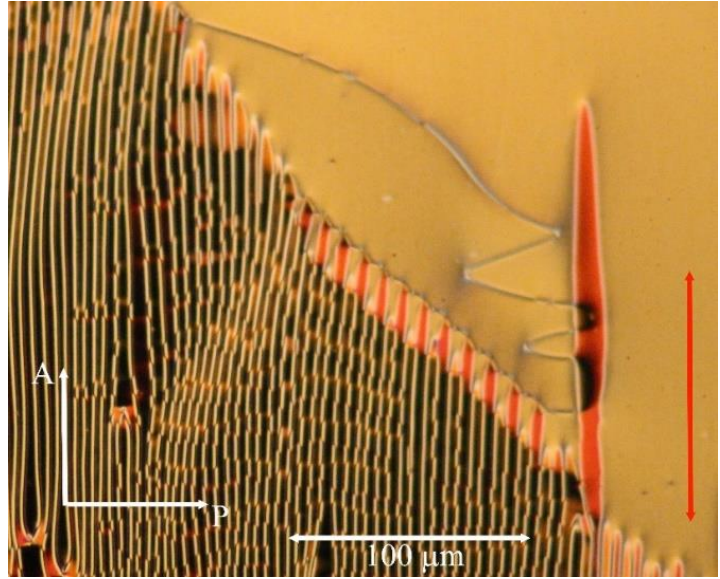

Figure 6.　　For FF6, the texture in the 5-μm HG-A cell, with the crossed polarizer (P) and analyser (A). In the right part there are twisted domains, the stripes in the left bottom part are oriented along the rubbing direction (red arrow).

As it was mentioned above, textural behaviour in HG cells with antiparallel rubbing shows a narrow temperature interval below the N phase, in which the texture looks homogeneous and the textural transformation to twisted domains appears at lower temperatures. Thus, we assumed the possibility of an intermediate phase between the N and $N_F$ phase. To resolve the presence of an additional phase, we performed high resolution AC calorimetry measurements in the vicinity of N-$N_F$ phase transition. Prior to measurements, the sample was heated twice to 154 °C, in the N phase, for 0.5 h. Then a cooling run was performed, with a rate of 150 mK/h, through the temperature range where the N- $N_F$ transition was expected from DSC. This rate is sufficiently slow to observe phase transitions with small enthalpy content, as it was recently demonstrated in the case of an intermediate phase between the N and the $N_F$ phases of RM734.[30] The temperature profile of the heat capacity, $c_p(T)$, is presented in Figure 7 and the anomaly consists of a single peak, indicating a direct transition between the N and $N_F$ phase.

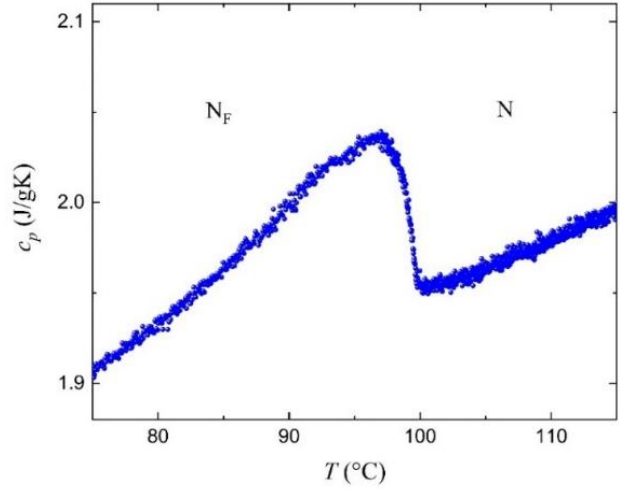

Figure 7.    The temperature profile of the heat capacity, $c_p(T)$, for the compound FF6 is shown upon cooling. The data were obtained by AC calorimetry using a scanning rate of 150 mK/h.

Additionally, we measured optical birefringence as a function of temperature for both compounds FF6 and ClF6 using the green light (550 nm). The birefringence, $\Delta n$, strongly jump-up at the Iso-N phase transition and follows a slight continuous increase within the N phase during the cooling process (see Figure 8 for FF6), reaching the values up to $\Delta n$ ~0.2. At the N-$N_F$ phase transition, there is an anomaly (inset in Figure 8), which evidences strong pre-transitional effects leading to a local decrease of birefringence. There is no evidence of an additional phase transition. Such an anomaly of $\Delta n(T)$ has already been described for other materials [46], for which it has been attributed to strong splay-type fluctuations at the N-$N_F$ phase transition. As there is only one distinct anomaly in the $\Delta n$ temperature dependence, we concluded that there is a direct phase transition between N and $N_F$ phase.

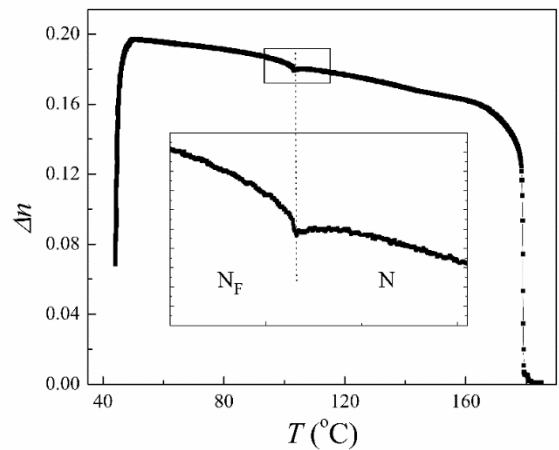

Figure 8.    Temperature dependence of the birefringence, $\Delta n$, at a wavelength of about 550 nm for compound FF6 measured in a 3-μm HG cell with parallel rubbing.

Dielectric spectroscopy has been performed in a broad range of frequencies from 1Hz to 1 MHz, in small measuring field of 0.01 V/μm. To eliminate a disturbing effect of surfactant layers, all measurements were performed on 5-μm home-made cells without surfactant. For permittivity, we detected a weak mode in the nematic phase, which grows for two orders of magnitude when approaching the $N_F$ phase on cooling. Three dimensional graphs of the real, ε', and imaginary part, ε'', of the complex permittivity for compound FF6 are presented in Figure 9, and for ClF6 in Figure S6 in SI. For FF6, we found a local anomaly at the N-$N_F$ phase transition temperature and in Figure S7 in SI, one can observe it in an enlarged view. We have fitted the real and imaginary parts of the permittivity and the dielectric strength, Δε, and relaxation frequencies, $f_r$, are shown Figure 10 for FF6 and in Figure S8 for the compound ClF6. The mode is strong and its dielectric strength in the $N_F$ phase exceeds $5\times10^4$. Far below the N-$N_F$ phase transition, the relaxation frequency decreases with the temperature, which is caused by an increase of the viscosity. A continuous phenomenological model of the $N_F$ phase developed by Vaupotic et al.[16] predicted a phason-type of the mode, which involves fluctuations of the molecular director. For compounds FF6 and ClF6, we confirmed that such a mode can be suppressed by a relatively small bias field. It is demonstrated in Figure S9, where we compared the real and imaginary parts of the permittivity without bias and under a bias from 0 to 0.6 V/μm.

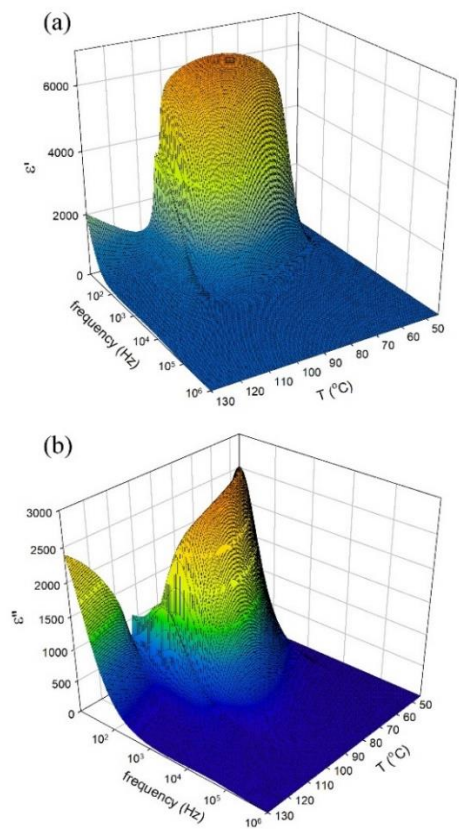

Figure 9. Permittivity versus temperature, *T*, and frequency in 3D-plots: (a) the real, ε', and (b) the imaginary, ε'', parts of permittivity for the compound FF6. Dielectric measurements were performed in a 5-μm cell without surfactant layers.

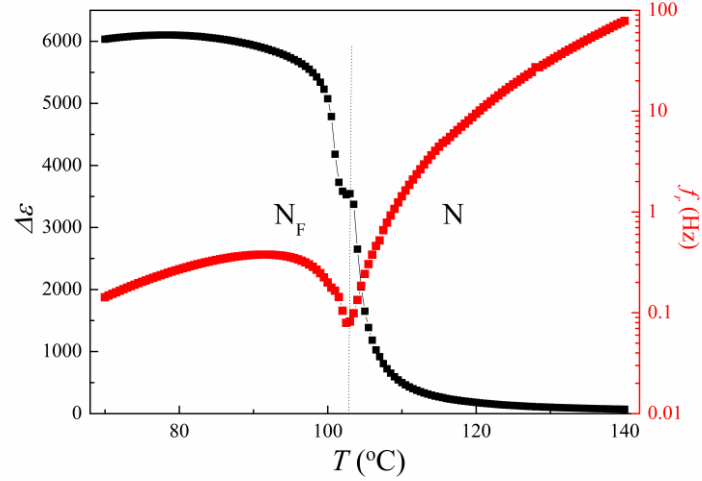

Figure 10.  Temperature dependence of the dielectric strength, $\Delta\varepsilon$, and the relaxation frequency, $f_r$, for compound FF6. Dielectric spectroscopy was performed on a 5-μm home-made cell without surfactant.

For both studied compounds FF6 and ClF6, we pursued switching properties in the ferroelectric phase and established the values of polarization, *P*. We applied 25-μm home-made cells without surfactant and detected a polarization current at a frequency of 20 Hz, under applied voltage ±1V/μm. From the current, we calculated polarization values and plotted temperature dependences *P*(T). In Figure 11, the temperature dependences are shown for both compounds in a relative temperature scale shifted towards the N-N$_F$ phase transition temperature, $T_c$. The *P* values grow continuously within the N$_F$ phase on the cooling bellow $T_c$, reaching up to 2.5 μC/cm$^2$ for ClF6 and 3 μC/cm$^2$ for FF6. In Supplementary file in Figure S10, a polarization current is shown for FF6 at the temperature T=85°C.

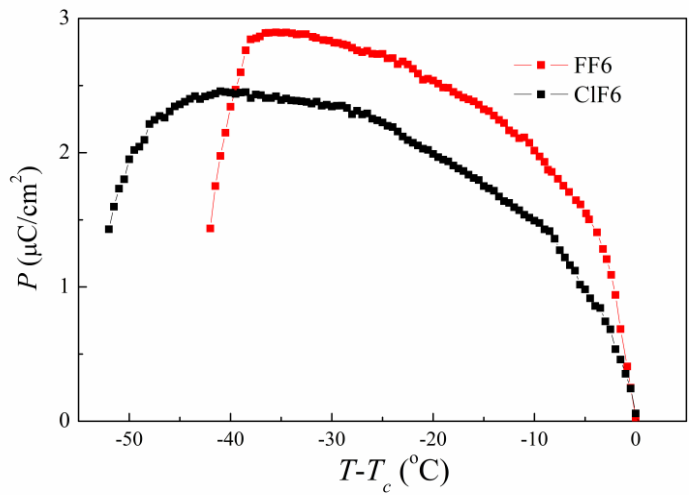

Figure 11.  Temperature dependence of polarization, *P*, for the compounds FF6 (red colour) and ClF6 (black) in a normalized temperature scale with respect to the N-N$_F$ phase transition temperature, $T_c$.

To confirm the absence of centrosymmetric properties of the $N_F$ phase, we conducted the SHG measurements in dependence on temperature (Figure 12). For both compounds, no SHG signal was observed above the N-$N_F$ phase transition temperature. Below the N-$N_F$ phase transition temperature, an abrupt increase appeared. For compound FF6, there is a narrow temperature interval below the N-$N_F$ phase transition, in which the SHG signal possessed anomalous behaviour. There is a local anomaly, which can be explained with a strong polar anchoring at the surfaces, which plays an important role at the vicinity of the N-$N_F$ phase transition even in the case of parallelly rubbed surfaces. Such effect in consistent with our textural observations described above.

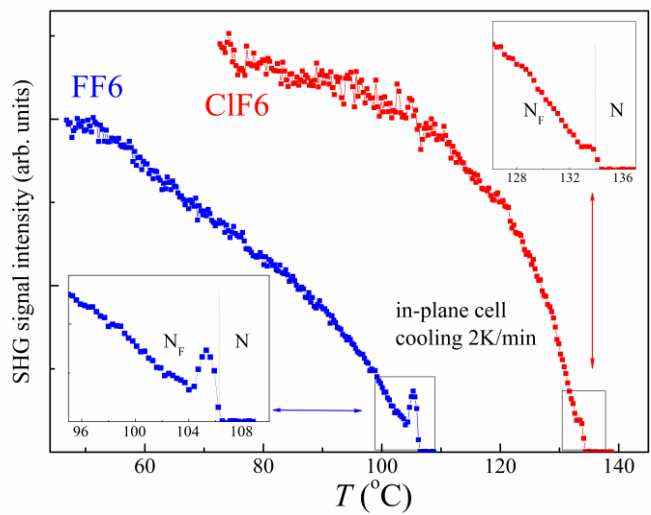

Figure 12. SHG signals for FF6 and ClF6 compounds in a planar 10-μm cell with parallel rubbing.

## 4 Discussion and conclusions

In this contribution, we present an innovative approach to the molecular structure of highly polar molecules exhibiting ferroelectric nematic phase. We have investigated the effects of the replacement of the usually applied electron donating groups by halogen atoms in the designed molecules. In the field of ferroelectric nematics, a chlorine atom has not yet been applied. Going from a small and highly electronegative fluorine to a bigger, less electronegative chlorine atom, with ca. 30 % longer bond, an increase of the transition temperatures has been observed. Nevertheless, such an unusual variation of the molecular structure represents a new contribution to the pool of ferroelectric nematogens, which is still rather limited to several molecular motives.

We synthesized three types of molecular structures and found that while three-ring F6 compound did not exhibit a mesomorphic behaviour, two four-ring compounds FF6 and ClF6 with prolonged molecular core revealed the N-$N_F$ phase sequence on cooling from the isotropic phase. Ferroelectricity in the $N_F$ phase has been proven by several experimental techniques including observation of textures, polarization measurements and the dielectric spectroscopy studies. Additionally, SHG measurements were accomplished to prove the ferroelectric behaviour and absence of symmetry centre in the $N_F$ phase.

The existence of the polar nematic phase for FF6 and ClF6 is consistent with the usual parameters of the formation of polar nematic phase described in the literature.[13,44] Based on the molecular calculations, new materials possess an adequate length, aspect ratio and a dipole moment at reasonable angle with respect to their long molecular axis. The calculated dipole values are: 7.79 D for F6, 10.58 D for FF6 and 9.94 D for ClF6. Only for two longer materials (FF6 and ClF6) the dipole exceeds the 9D threshold, which was established to be necessary for the formation of polar phases.[13] The C-Cl bond dipole is oriented antiparallelly to the EWG part of the molecule and possesses higher value than C-F bond dipole. Due to this reason, the material terminated with chlorine atom has slightly lower total dipole moment and the overall polarization values (Figure 11).

We have observed a strong tendency to homeotropic orientation in the N phase. While in paraelectric nematic phase (N) we got homeotropic texture even without any surface anchoring. Nevertheless, such a texture disappeared after the transition to the $N_F$ phase and the texture changed to a more complex one. As a uniform homeotropic alignment in a polar phase in highly improbable unless under electric field, the transformation to a schlieren texture is natural. Nevertheless, it is still mysterious why homeotropic texture in the N phase is preferentially realized under ITO electrodes and what character has the anchoring at different surfaces. Due to the anchoring and surface effects, we have detected unusual textural transformation in the $N_F$ phase in planar cells. The twisted state in the $N_F$ phase did not appear immediately below the N-$N_F$ phase transition. There is a narrow temperature interval in the $N_F$ phase, in which the textures reveal paramorphic character, resembling features from the upper N phase. Nevertheless, this state is unstable, and finally twisted states appear in a form of anchoring transition. Recently, similar behaviour has been described for other ferroelectric nematogens.[48]

As textures in polarized light shows strong pretransition effects and specific textural transformation (Figure 2) in cells with antiparallel rubbing, we performed the AC calorimetric measurements to ascertain mesogenic properties in a bulk. In this experiment no additional anomalies were found at the vicinity of the N-$N_F$ phase transition, contrary to POM observations and SHG results. As AC calorimetry technique is carried out on a thick sample, without any anchoring conditions imposed, these results serve as a reliable proof of the phase transition identification. Additionally, precise birefringence measurements confirmed the direct N- $N_F$ phase sequence and excluded any intermediate phase. Both experiments show rather broad phase transition region between the nematic and the ferroelectric nematic phase, which is usually attributed to strong splay-type fluctuations close to the transition point.

We can conclude that presented herein new compounds possess a broad ferroelectric nematic phase with excellent properties, concurring with previously reported variants of ferroelectric nematogens. The presented results can be beneficial for further development in the field of soft matter, as ferroelectric nematogens have strong potential of technological applications.

**Acknowledgements**
Authors acknowledge the financial support of the Czech Science Foundation, the project 24-10247K (the Czech Science Foundation) and the Slovenian Research and Innovation Agency (ARIS) programme P1-0125. V.N. is grateful to Damian Pociecha from Warsaw University for fruitful discussion. M.L. is thankful to Zdravko Kutnjak for allowing access to AC calorimetry and to George Cordoyiannis for useful comments.

# The effect of fluorine or chlorine substitution on mesomorphic properties of ferroelectric nematic liquid crystals

Martin Cigl,[1] Natalia Podoliak,[1] Dalibor Repček,[1] Pavlo Golub,[1] Marta Lavrič,[2] and Vladimíra Novotná [1]*

[1] *Institute of Physics of the Czech Academy of Sciences, Na Slovance 1999/2, 182 00 Prague 8, Czech Republic*

[2] *Jožef Stefan Institute, Jamova cesta 39, SI-1000 Ljubljana, Slovenia*

**Content**

1. **Synthesis of materials**
   1.1. General
   1.2. Synthetic procedures
2. **Experimental set-ups and methods**
3. **Results**

1. **Synthesis of materials**

1.1. **General**

All starting materials and reagents were purchased from Sigma-Aldrich, Acros Organics or Lach:Ner with purity "For synthesis" or better. All solvents used for the synthesis were "p.a." grade. $^1$H NMR spectra were recorded using Varian VNMRS 300 instrument, deuteriochloroform (CDCl$_3$) and hexadeuteriodimethyl sulfoxide (DMSO-$d_6$) were used as solvents and the signals of the solvents served as internal standards. Chemical shifts (d) are given in ppm and *J* values are given in Hz. The signals were verified with the help of homonuclear and heteronuclear correlation experiments. Elemental analyses were carried out on Elementar vario EL III instrument. The purity of all final compounds was checked by HPLC analysis (high-pressure pump ECOM Alpha; column WATREX Biospher Si 100, 250 × 4 mm, 5 mm; detector WATREX UVD 250) and was found to be >99.8 %. Column chromatography was carried out using Merck Kieselgel 60 (60-100 μm or 40-60 μm).

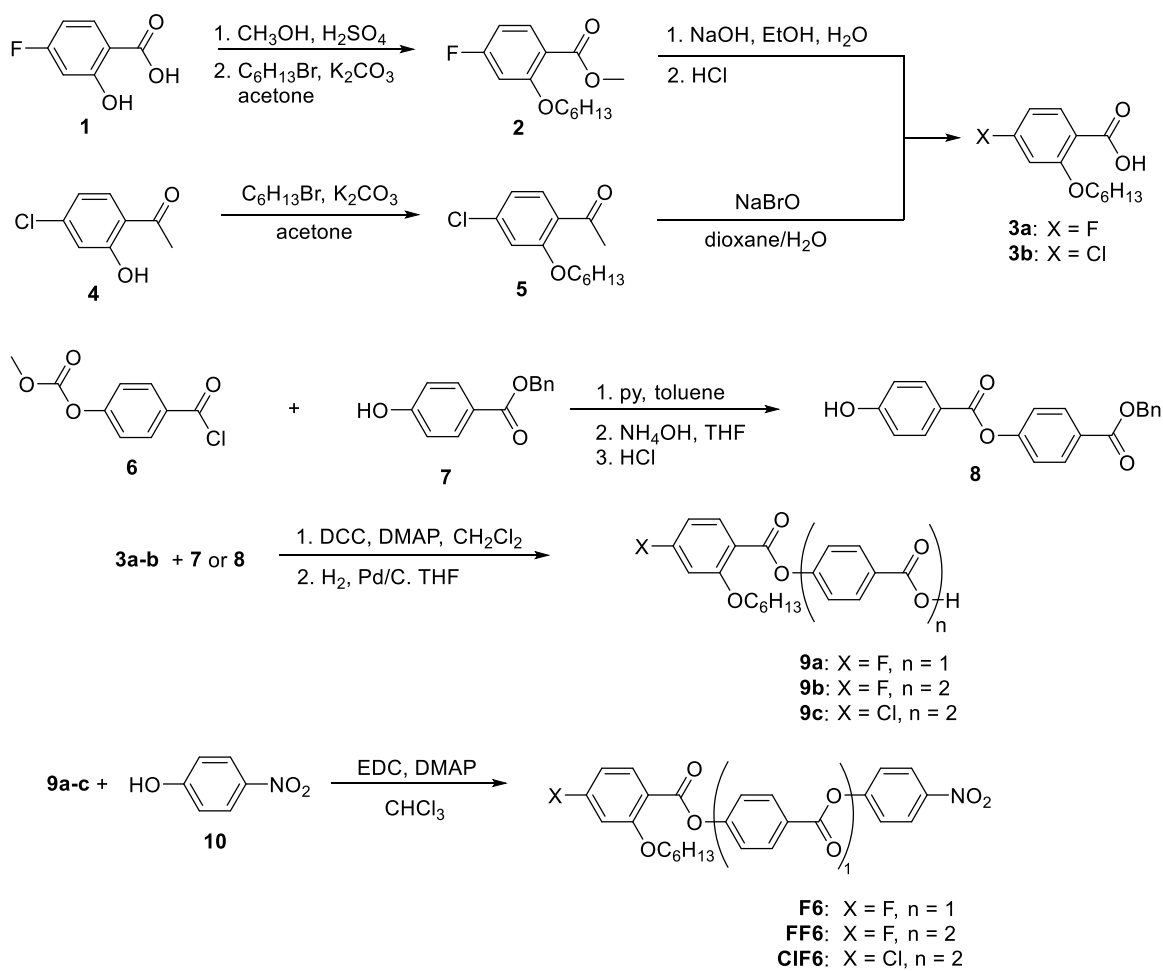

Scheme S1.    Synthetic route to target mesogens.

The synthesis of the fluoro-terminated material started from the commercially available 4-fluorosalicylic acid (**1**), which was converted to ester and alkylated with *n*-bromohexane. After the basic hydrolysis of the methylester group of **2**, the acid **3a** was obtained. The chlorinated analogue was synthesized in two steps from 4-chloro-2-hydroxyacetophenone (**4**). The first step was an introduction of the sidechain via alkylation reaction and the second comprised haloform reaction to transform the acetyl group to carboxylic giving acid **3b**. Next, the benzoyl chloride **6**, prepared as described in Ref. [S1], was used in the esterification reaction with benzyl 4-hydroxybenzoate (**7**) and the protective methoxycarbonyl group was subsequently removed by means of aqueous ammonia in THF to yield benzyl-protected intermediate **8**. The halogenated acids **3a-b** were combined with the hydroxybenzoate **7** or **8** in a DCC-mediated esterification reaction, after which the benzyl protective group was removed by the hydrogenolysis on palladium. The obtained intermediates **9a-c** were esterified with 4-nitrophenol (**10**) using EDC coupling to give target mesogens **F6**, **FF6** and **ClF6**.

## 1.2. Synthetic procedures

*Methyl 4-fluoro-2-(hexyloxy)benzoate* (**2**)

4-Fluoro-2-hydroxybenzoic acid (10.0 g, 62.14 mmol) was dissolved in methanol 120 mL and H$_2$SO$_4$ (conc. 5 mL) was added dropwise, and then the reaction mixture was refluxed for 4 h. The reaction volume was reduced to ca. 1/5, diluted with ethyl acetate (100 mL) and washed with water (3 × 30 mL). After drying over anhydrous MgSO$_4$, the solvent was removed and the crude ester was used in the next step without further purification. *n*-Bromohexane (10.35 g, 62.14 mmol) was added to methyl 2-fluoro-4-hydroxybenzoate, anhydrous potassium carbonate (16.0 g, 0.12 mol) and potassium iodide (0.10 g) in acetone (350 mL). The reaction mixture was stirred under reflux under anhydrous conditions (CaCl$_2$ tube) for 24 h. The cooled mixture was filtered, and the filtrate evaporated. The residue was diluted with diethylether (200 mL) and washed with NaOH (70 mL, 5%), water (70 mL) and dried over anhydrous MgSO$_4$. The solvent was removed under reduced pressure and the residue crystallized from hexane at low temperature (2 °C). Yield 13.96 g (88 %). $^1$H NMR (CDCl$_3$): 7.83 (1 H, dd, *J*=9.39, 7.04 Hz), 6.58 - 6.70 (2 H, m), 3.99 (2 H, t, *J*=6.46 Hz), 3.87 (3 H, s), 1.77 - 1.92 (2 H, m), 1.42 - 1.57 (2 H, m), 1.26 - 1.42 (4 H, m), 0.90 (3 H, t. *J*=6.7).

*4-Fluoro-2-(hexyloxy)benzoic acid* (**3a**)

Benzoate **2** (13.50 g, 53.09 mmol) was dissolved in a mixture of ethanol (100 mL) heated to ca 50 °C with stirring. Solution of NaOH (5.31 g, 0.13mol) in aqueous ethanol (100 mL 90 %) was added and the reaction mixture stirred at 50 °C for 2 h. The resulting mixture was placed on a rotavap to reduce the volume to ca ¼ and then poured into water (200 mL), carefully neutralized with HCl, and extracted with ethyl acetate (3 × 70 mL). Yield (12.37 g, 97 %). $^1$H NMR (CDCl$_3$): 8.52 (1 H, dd, *J*=9.39, 7.04 Hz), 6.71 - 6.82 (2 H, m), 4.19 (2 H, t, *J*=6.46 Hz), 1.86 - 1.95 (2 H, m), 1.33 - 1.48 (6 H, m), 0.93 (3 H, t, *J*=6.7).

*4-Chloro-2-(hexyloxy)acetophenone* (**5**)

*n*-Bromohexane (3.84 g, 22.80 mmol) was added to a stirred mixture of 4-chloro-2-hydroxyacetophenone (**4**, 2.0 g, 11.72 mmol), potassium iodide (0.10 g) and anhydrous potassium carbonate (3.00 g, 21.71 mmol) in acetone (120 mL) and the reaction mixture was refluxed under anhydrous conditions for 16 h. After cooling, the resulting mixture was filtered and the filtrate evaporated. The residue was dissolved in ethyl acetate, washed with saturated NaHCO$_3$, brine and dried with MgSO$_4$. Removal of the solvent on evaporator yielded product **5** (3.13 g, 97 %, colourless liquid), of 95 % purity, which was used in the next step without further purification. $^1$H NMR (CDCl$_3$): 7.69 (1 H, d, *J*= 7.50 Hz), 6.90 - 6.95 (2 H, m), 4.01 (2 H, t, *J*=6.46 Hz), 1.58 (3 H, s). 1.79 - 1.90 (2 H, m), 1.23 - 1.47 (6 H, m), 0.89 (3 H, t, *J*=6.7).

*4-Chloro-2-(hexyloxy)benzoic acid* (**3b**)

Sodium hypobromite solution was prepared by dropwise addition of bromine (4.0 mL, 124.25 mmol) to the solution of sodium hydroxide (4.97 g, 124.25 mmol) in 25 mL water at 0 °C and added dropwise to the stirred solution of 2-chloro-4-(hexyloxy)acetophenone (2.66 g, 9.93 mmol) in dioxane (50 mL) with such a rate that the reaction temperature is not higher than 40 °C (ca. 15 min). The reaction mixture was stirred at 25 °C overnight. The excess of hypobromite was eliminated by the addition of sodium bisulphite and the reaction mixture acidified with HCl to pH = 2. The resulting mixture was extracted with diethylether (3×), the combined organic layers were washed with water, brine and dried with MgSO$_4$. After removal of the solvent, the crude product was crystallised from hexane. Yield 2.32 g (9.07 mmol, 91 %) acid **3b**, m.p. 82 – 83 °C. $^1$H NMR (CDCl$_3$): 10.71 (2 H, br. s.), 8.12 (1 H, d, *J*=8.2 Hz), 7.12 (1 H, dd, *J*=8.2, 1.8 Hz), 7.04 (1 H, d, *J*=1.8 Hz), 4.24 (2 H, t, *J*=6.5 Hz), 1.92 (2 H, quin, *J*=7.2 Hz), 1.28 - 1.56 (6 H, m), 0.91 (3 H, t, *J*=6.7 Hz).

*Benzyl 4-((4-hydroxybenzoyl)oxy)benzoate* (**8**)
Solution of the benzoyl chloride **6** prepared from the corresponding benzoic acid (10.0 g, 51.0 mmol) as described in Ref. [S1] in toluene (100 mL) at ca 60 °C was poured into the stirred mixture of benzyl 4-hydroxybenzoate (**7**, 11.64 g, 51.0 mmol) and pyridine (4.10 mL) in toluene (100 mL). The reaction mixture was stirred for ca 6 h under anhydrous conditions (CaCl$_2$ tube). Water (100 mL) acidified with HCl (10 mL, 35 %) was added to the resulting mixture. The organic layer was separated, washed with saturated solution of NaHCO$_3$ (50 mL) and water (50 mL) and dried with anhydrous MgSO$_4$. The solvent was removed under reduced pressure and the crude intermediate crystallized from toluene to obtain the first portion of the product. The second portion was obtained by maceration of the evaporated mother liquor with diethyl ether, resulting in total yield of 19.55 g.

The obtained intermediate was dissolved in THF (120 mL) and cooled to ca. -10 °C in an ice-water-salt bath. To this solution, aqueous ammonium hydroxide (10 mL, 25%) was added in one portion. The reaction was let warm to 5 °C and the progress of hydrolysis was monitored by TLC (CH$_2$Cl$_2$ : acetone, 97 : 3). After ca. 2 h, the reaction mixture was neutralized with HCl and extracted with ethyl acetate (3 × 60 mL). Combined organic layers were washed with water, brine and dried with anhydrous MgSO$_4$. After removal of the solvent, the crude product was crystallized and the residue from mother liquor was further purified by column chromatography on silica (CH$_2$Cl$_2$ : acetone, 97 : 3, R$_f$ = 0.3). Yield 10.11 g (85 %). $^1$H NMR (CDCl$_3$): 8.16 (2 H, d, *J*=8.8 Hz), 8.10 (2 H, d, *J*=8.8 Hz), 7.33 - 7.54 (5 H, m), 7.29 (2 H, d, *J*=8.8 Hz), 6.91 (2 H, d, *J*=8.8 Hz), 5.39 (2 H, s).

*4-((4-Fluoro-2-(hexyloxy)benzoyl)oxy)benzoic acid* (**9a**)
Benzoic acid **3a** (3.0 g, 12.49 mmol) and hydroxybenzoate **7** (2.85 g, 12.49 mmol) were dissolved in CH$_2$Cl$_2$ (50 mL) and cooled to ca. 10 °C in ice-water bath and 4-(*N,N*-dimethylamino)pyridine (DMAP, 0.75 g, 5.83 mmol) was added. Then *N,N′*-dicyclohexylcarbodiimide (DCC, 5.15 g, 24.71 mmol) and the reaction mixture was stirred under anhydrous conditions (CaCl$_2$ tube) for 3 h. The precipitated *N,N′*-dicyclohexylurea was filtered off and the filtrate evaporated on a rotary evaporator. Crude product was then purified by column chromatography on silica gel (CH$_2$Cl$_2$ : acetone, 98 : 2, R$_f$ = 0.7) and crystallized from hexane.

The obtained intermediate (5.03 g) was dissolved in THF (300 mL) and palladium on charcoal (0.50 g, 10%, unreduced) was added. The reaction flask was evacuated and refilled with hydrogen from a balloon. Reaction was stirred for 30 minutes and then evacuated again and the solid was filtered off. Filtrate was evaporated and the residue crystallized from ethanol to yield pure benzoic acid **9a**. Yield 3.68 g (82 %). $^1$H NMR (CDCl$_3$): 8.20 (2 H, d, *J*=8.8 Hz), 8.05 (1 H, dd, *J*=9.1, 6.7 Hz), 7.33 (2 H, d, *J*=8.8 Hz), 6.66 - 6.81 (2 H, m), 4.05 (2 H, t, *J*=6.2

Hz), 1.77 - 1.95 (2 H, m), 1.41 - 1.57 (2 H, m), 1.21 - 1.39 (4 H, m), 0.86 (3 H, t, *J*=6.7 Hz). $^{13}$C {$^{1}$H} NMR (CDCl$_3$): 171.52 (s), 166.73 (d *J*=255.49), 163.05 (s), 161.64 (d, *J*=10.4 Hz), 155.39 (s), 134.61 (s), 134.47 (s), 131.85 (s), 126.70 (s), 121.93 (s), 114.66 (d, *J*=3.32 Hz), 107.15 (d, *J*=21.96 Hz), 100.95 (d, *J*=25.44 Hz), 69.30 (s), 31.41 (s), 28.89 (s), 25.58 (s), 22.50 (s), 13.93 (s).

*4-((4-((4-Fluoro-2-(hexyloxy)benzoyl)oxy)benzoyl)oxy)benzoic acid* (**9b**)
Following the procedure for the compound **9a**: Benzoic acid **3a** (1.10 g, 4.55 mmol) was esterified with phenol **8** (1.59 g, 4.55 mmol) using DCC (1.02 g, 5.0 mmol) and DMAP (0.27 g, 2.23 mmol). Obtained benzoate (2.30 g, 4.79 mmol) was hydrogenolysed using Pd/C in THF and boiled in heptane to yield 1.81 g (83 %) $^{1}$H NMR (CDCl$_3$): 8.29 (1 H, d, *J*=8.8 Hz), 8.22 (2 H, d, *J*=8.8 Hz), 8.07 (1 H, dd, *J*=9.1, 6.7 Hz), 7.31 - 7.43 (3 H, m), 6.68 - 6.83 (2 H, m), 4.07 (2 H, t, *J*=6.5 Hz), 1.78 - 1.95 (2 H, m), 1.42 - 1.59 (2 H, m), 1.20 - 1.42 (4 H, m), 0.87 (3 H, t, *J*=6.7 Hz). $^{13}$C {$^{1}$H} NMR (CDCl$_3$): 170.92 (s), 166.82 (d, *J*=255.48 Hz), 163.85 (s), 163.02 (s), 161.71 (d, *J*=10.4 Hz), 155.51 (s), 155.26 (s), 134.68 (s), 134.53 (s), 131.95 (s), 131.91 (s), 126.89 (s), 126.36 (s), 122.20 (s), 121.93 (s), 114.59 (d, *J*=3.5 Hz), 107.23 (d, *J*=21.96 Hz), 101.01 (d, *J*=25.43 Hz), 69.34 (s), 31.44 (s), 28.92 (s), 25.61 (s), 22.53 (s), 13.97 (s).

*4-((4-((4-Chloro-2-(hexyloxy)benzoyl)oxy)benzoyl)oxy)benzoic acid* (**9c**)
Using the procedure described for **9a**: Benzoic acid **3b** (0.44 g, 1.72 mmol) was reacted with benzoate **8** (0.50 g, 1.44 mmol) in the presence of DCC (0.34 g, 1.60 mmol) and DMAP (0.10 g, 0.82 mmol) in 10 mL of CH$_2$Cl$_2$ to yield 0.92 g of benzylated intermediate, which was hydrogenolysed in ethyl acetate (100 mL) using Pd/C (48.0 mg). Yield 0.67 g (78 %). $^{1}$H NMR (CDCl$_3$): 11.20 (1 H, br. s.), 8.28 (2 H, d, *J*=8.8 Hz), 8.22 (2 H, d, *J*=8.8 Hz), 7.97 (1 H, d, *J*=8.8 Hz), 7.29 - 7.46 (4 H, m), 7.03 (2 H, dd, *J*=4.4, 2.6 Hz), 4.07 (2 H, t, *J*=6.5 Hz), 1.78 - 1.93 (2 H, m), 1.50 (2 H, quin, *J*=7.3 Hz), 1.21 - 1.41 (4 H, m), 0.82 - 0.92 (3 H, m). $^{13}$C {$^{1}$H} NMR (CDCl$_3$): 171.29 (s), 163.79 (s), 163.16 (s), 160.16 (s), 155.40 (s), 155.23 (s), 140.64 (s), 133.41 (s), 131.93 (s), 131.89 (s), 126.90 (s), 126.40 (s), 122.13 (s), 121.90 (s), 120.38 (s), 116.93 (s), 113.71 (s), 69.31 (s), 31.41 (s), 28.95 (s), 25.57 (s), 22.51 (s), 13.96 (s).

*4-((4-Nitrophenoxy)carbonyl)phenyl 4-fluoro-2-(hexyloxy)benzoate* (**F6**)
Benzoic acid **9a** (4.10 g, 11.37 mmol) and 4-nitrophenol (**10**, 1.58 g, 11.36 mmol) were dissolved in dry dichloromethane (80 ml) and cooled to 10 °C in an ice-water bath. Then *N*-(3-dimethylaminopropyl)-*N*′-ethylcarbodiimide hydrochloride (EDC, 2.40 g, 12.45 mmol) and 4-(*N*,*N*-dimethylamino)pyridine (DMAP, 0.69 g, 5.37 mmol) were added. The reaction mixture was stirred for 2 hours under anhydrous conditions and the temperature let rise as ice in the cooling bath melted. The volume of the resulting mixture was reduced on rotary evaporator and the residue (ca 5 mL) filtered through a short silica gel column (100 mL, diameter 70 mm) and eluted with CH$_2$Cl$_2$-acetone (95:5). Collected fraction with the product was evaporated and the residue was purified by column chromatography on silica gel in CH$_2$Cl$_2$-acetone 99 : 1 eluent and recrystallized from acetonitrile. Yield 2.66 g (49 %). $^{1}$H NMR (CDCl$_3$): 8.34 (2 H, d, *J*=9.4 Hz), 8.28 (2 H, d, *J*=8.8 Hz), 8.01 - 8.13 (1 H, m), 7.41 (4 H, dd, *J*=12.3, 8.8 Hz), 6.68 - 6.86 (2 H, m), 4.06 (2 H, t, *J*=6.5 Hz), 1.78 - 1.94 (2 H, m), 1.41

- 1.61 (3 H, m), 1.17 - 1.41 (4 H, m), 0.87 (3 H, t, $J$=6.7). $^{13}$C {$^{1}$H} NMR (CDCl$_3$): 166.83 (d, $J$=255.48 Hz), 163.51 (s), 162.91 (s), 161.73 (d, $J$=10.4 Hz), 155.69 (s), 155.60 (s), 145.40 (s), 134.62 (d, $J$=11.6 Hz), 131.97 (s), 125.79 (s), 125.28 (s), 122.62 (s), 122.32 (s), 114.35 (s), 109.96 (s), 107.23 (d, $J$=21.97 Hz), 100.96 (d, $J$=25.43 Hz), 69.30 (s), 31.42 (s), 28.89 (s), 25.58 (s), 22.51 (s), 13.97 (s). Anal. calcd. for C$_{26}$H$_{24}$FNO$_7$: C 64.86, H 3.95, N 2.91; found C 64.74, H 4.74, N 2.88 %.

*4-((4-((4-Nitrophenoxy)carbonyl)phenoxy)carbonyl)phenyl 4-fluoro-2-(hexyloxy)benzoate* (**FF6**)

Benzoic acid **9b** (1.53 g, 3.18 mmol) was esterified with 4-nitrophenol (**10**, 0.89 g, 6.33 mmol) using EDC (0.68 g, 3.49 mmol) and DMAP (0.20 g, 1.58 mmol) in dichloromethane (7.0 mL) analogously as described for **F6**. Chromatography: CH$_2$Cl$_2$-acetone 98 : 2, Crystallisation from acetonitrile. Yield 1.06 g (55 %). $^1$H NMR (CDCl$_3$): 8.22 - 8.44 (6 H, m), 8.01 - 8.14 (1 H, m), 7.32 - 7.52 (6 H, m), 6.74 (2 H, d, $J$=10.6 Hz), 4.07 (2 H, t, $J$=6.5 Hz), 1.77 - 1.96 (2 H, m), 1.43 - 1.61 (3 H, m), 1.17 - 1.42 (4 H, m), 0.87 (3 H, t, $J$=6.7). $^{13}$C {$^{1}$H} NMR (CDCl$_3$): 166.82 (d, $J$=255.48 Hz) 163.79 (s) 163.46 (s) 162.96 (s) 161.72 (d, $J$=11.6 Hz) 155.56 (s) 145.44 (s) 134.61 (d, $J$=11.6 Hz) 132.07 (s) 131.93 (s) 126.16 (s) 126.09 (s) 125.30 (s) 122.64 (s) 122.27 (s) 114.45 (s) 107.22 (d, $J$=21.96 Hz) 100.97 (d, $J$=25.43 Hz) 69.30 (s) 31.43 (s) 28.91 (s) 25.60 (s) 22.53 (s) 13.99 (s). Anal. calcd. for C$_{33}$H$_{28}$FNO$_9$: Anal. calcd. for C$_{23}$H$_{20}$N$_2$O$_7$: C 65.89, H 4.69, N 2.33; found C 66.46, H 4.75, N 2.34 %.

*4-((4-((4-Nitrophenoxy)carbonyl)phenoxy)carbonyl)phenyl 4-chloro-2-(hexyloxy)benzoate* (**ClF6**)

Benzoic acid **9c** (1.60 g, 3.22 mmol) was esterified with 4-nitrophenol (**10**, 0.46 g, 3.27 mmol) using EDC (0.69 g, 3.51 mmol) and DMAP (0.20 g, 1.58 mmol) in dichloromethane (15.0 mL) analogously, as described for **F6**. Chromatography: CH$_2$Cl$_2$-acetone 99 : 1, Crystallisation from acetonitrile. Yield 1.01 g (50 %). $^1$H NMR (CDCl$_3$): 8.24 - 8.41 (6 H, m), 7.98 (1 H, d, $J$=8.8 Hz), 7.35 - 7.50 (6 H, m), 6.99 - 7.09 (2 H, m), 4.08 (2 H, t, $J$=6.5 Hz), 1.78 - 1.95 (2 H, m), 1.42 - 1.55 (2 H, m), 1.22 - 1.40 (4 H, m), 0.87 (3 H, t, $J$=6.7). $^{13}$C {$^{1}$H} NMR (CDCl$_3$): 168.51 (s), 165.12 (s), 163.79 (s), 163.45 (s), 162.96 (s), 161.79 (s), 161.64 (s), 155.55 (s), 145.43 (s), 134.68 (s), 134.53 (s), 132.06 (s), 131.92 (s), 126.16 (s), 126.08 (s), 125.30 (s), 122.63 (s), 122.26 (s), 114.44 (s), 107.36 (s), 107.07 (s), 101.13 (s), 100.79 (s), 69.30 (s), 31.42 (s), 28.91 (s), 25.59 (s), 22.53 (s), 13.99 (s). Anal. calcd. for C$_{33}$H$_{28}$ClNO$_9$: C 63.95, H 5.74, N 2.27; found C 63.95, H 4.51, N 2.25 %.

## 2. Experimental procedures and methods

For textural observations, various commercial cells have been purchased from WAT company, Poland. Cells with homogeneous geometry (HG) were utilized, having the rubbing direction of the surfactant layer parallel (HG-P) or antiparallel (HG-A) at opposite surfaces. For comparison, the cell with homeotropic anchoring (HT) was applied, in a geometry with molecules perpendicular to the cell surface. For the dielectric spectroscopy and polarization measurements, home-made 5-µm sandwich-type glass cells without surfactant were prepared. For the second harmonic generation measurements, HG-P cells of thickness 9-10 µm were used. All the cells were filled with the studied materials in the isotropic phase by means of capillary action. For the polarized light microscopy (POM) investigations, Nikon Eclipse E600 microscope equipped with a photo camera Canon EOS 700D was applied. A heating-cooling stage with an accuracy ±0.1 K and a temperature controller Linkam TMS 94 (Linkam, Tadworth, UK) was used for temperature stabilization.

To establish the basic information about all phase transition temperatures and enthalpies, differential scanning calorimetry (DSC) measurements were performed using calorimeter Pyris 7 (Perkin Elmer, Shelton, CT, USA). The samples of 2-5mg were hermetically sealed into aluminium pans and placed into the calorimeter chamber infiltrated with nitrogen. The samples were measured during heating and cooling cycles at a rate of 10 K/min. The temperatures and enthalpies were evaluated from endothermic and exothermic peaks.

For high-resolution AC calorimetric measurements, a sample of mass of 30 mg was loaded into a high-purity, indium-sealed silver cell. A heater and small glass bead thermistor were attached on opposite sides of the cell to apply an oscillating power on the sample and measure the changes in its temperature, respectively. The usual AC mode of operation senses the continuous changes of enthalpy and yields the precise temperature profile of the specific heat capacity $c_p(T)$ in the case of second-order transitions. This mode does not sense a latent heat, however, a first-order transition can be qualitatively distinguished from the phase shift between the applied AC power and the temperature oscillations of the sample. An extensive description of the technique can be found in Ref. [S2]. Home-made experimental setup was developed at Jožef Stefan Institute, its thermal stability is better than 0.1 mK and it allows very slow cooling and heating runs. The heat capacity of the empty cell and other components (heater, thermistor) is subtracted, and the result is then divided by the sample mass to obtain the net specific heat capacity of the sample.

Dielectric spectroscopy studies were conducted with Schlumberger 1260 impedance analyser (Schlumberger, Houston, TX, USA). The sample temperature was stabilised within ±0.1 K during the frequency sweeps (1 Hz ÷ 1 MHz). From the resistance and capacity of the sample, real and imaginary parts of the complex permittivity, $\varepsilon^*(f) = \varepsilon' - i\varepsilon''$, were obtained. The data were subsequently fitted to Cole-Cole equation:

$$\varepsilon^* - \varepsilon_\infty = \frac{\Delta\varepsilon}{1+(if/f_r)^{(1-\alpha)}} - i\left(\frac{\sigma}{2\pi\varepsilon_0 f^n} + Af^m\right) \qquad (1)$$

where $f_r$ is the relaxation frequency, $\Delta\varepsilon$ is the dielectric strength, $\alpha$ is the distribution parameter of relaxation, $\varepsilon_0$ is the permittivity of vacuum, $\varepsilon_\infty$ is the high frequency permittivity, n, m, and A are the parameters of fitting. From this procedure, $\Delta\varepsilon$ and $f_r$ were acquired in dependence of temperature, $T$. Polarization, $P$, measurements were performed under the triangle-profile electric field with the frequency of 10 Hz and the magnitude of 10 V/μm, supplied by an Agilent generator (Agilent, California, US). A switching current at stabilized temperature was detected using Tektronix DPO4034 (Tektronix, Beaverton, USA) digital oscilloscope and was integrated to supply $P(T)$ values.

The temperature-dependent second harmonic generation (SHG) measurements were performed in transmission configuration with an optical set-up based on femtosecond Ti:sapphire laser (Spitfire ACE). The laser beam was amplified to produce 40 fs long pulses with 5 kHz repetition rate and the central wavelength of 800 nm. The rubbing direction of the HG-P cell was oriented parallel to the polarization direction of the incident beam. The samples were illuminated with a collimated beam with pulses fluence of approximately 0.01 mJ/cm$^2$. The generated SGH signal was spectrally filtered using optical dichroic mirrors with the central wavelength of 400 nm, detected with an avalanche photodiode and subsequently amplified using a lock-in amplifier.

Birefringence measurements were performed on thin 1.6-µm HG-P cells in a set-up equipped with a green light ($\lambda$ = 550 nm) source. The light beam passed through a polariser and a photoelastic modulator PEM-100 (Hinds instruments), a sample placed in Instec heating stage, controlled with Instec STC 200 controller, an analyser and a photodetector. The birefringence, $\Delta$n, was calculated from the detected values of the retardation.

## 3. Results

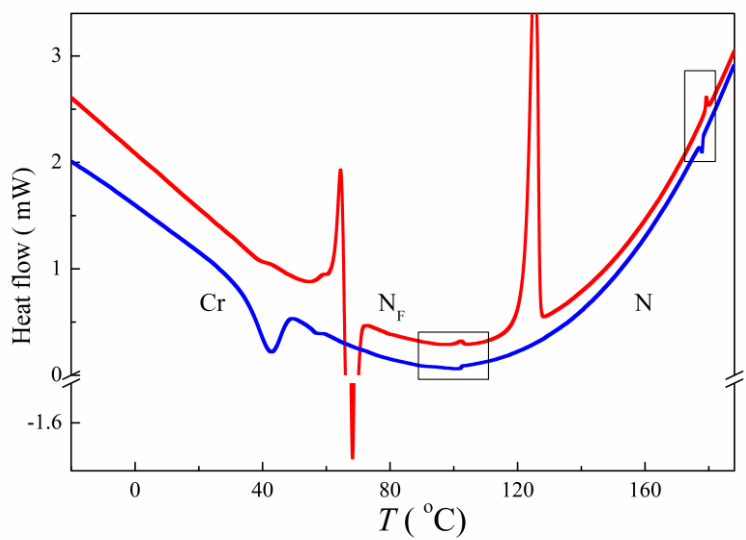

Figure S1.    For compound FF6, DSC thermographs detected during the second heating (red) and cooling (blue) runs. The boxes mark the areas of the phase transitions.

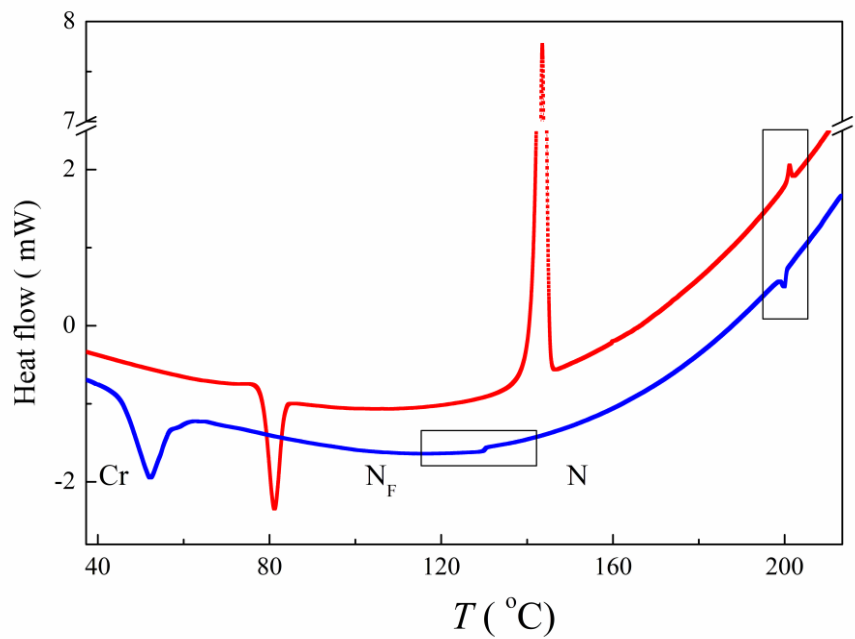

Figure S2. For compound ClF6, DSC thermographs detected during the second heating (red) and cooling (blue) runs. The boxes mark the areas of the phase transitions.

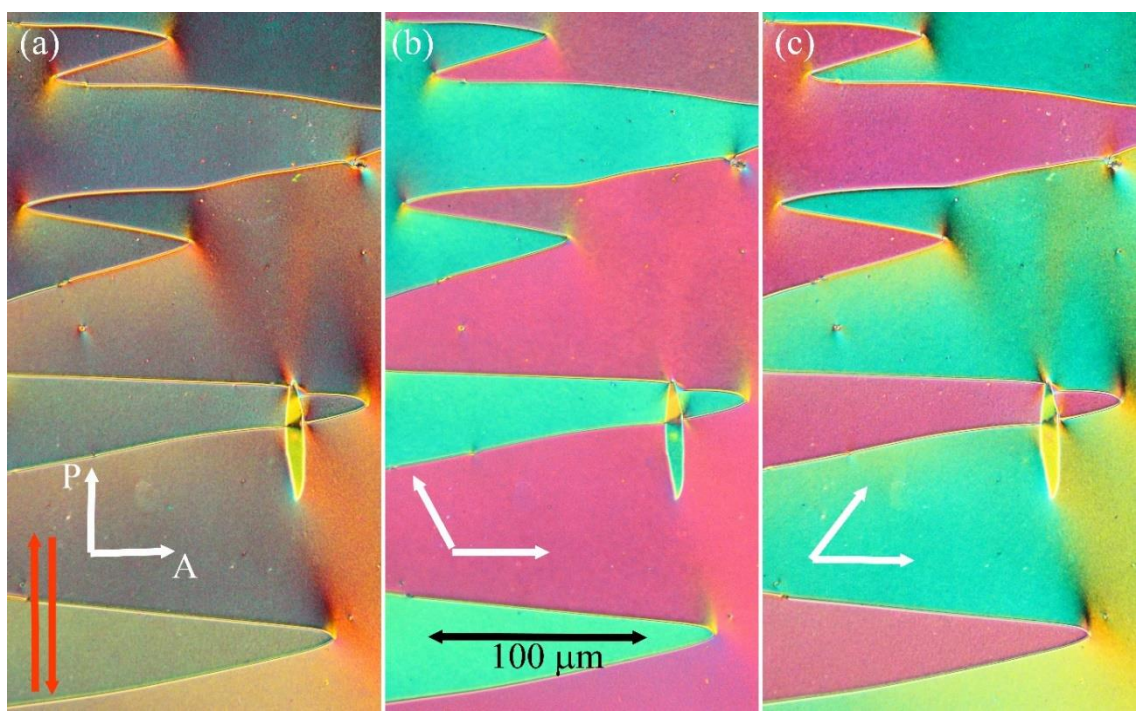

Figure S3. FF6 in a 5-μm HG-A cell with antiparallel rubbing (red arrows) in the $N_F$ phase at T=40°C, (a) in the crossed position of polarizers; (b) and (c) show the same view in uncrossed position, when the polarizer (P) is rotated at an angle of about 20 degrees anticlockwise or clockwise from the analyser (A). Boarder between domains corresponds to a disclination line.

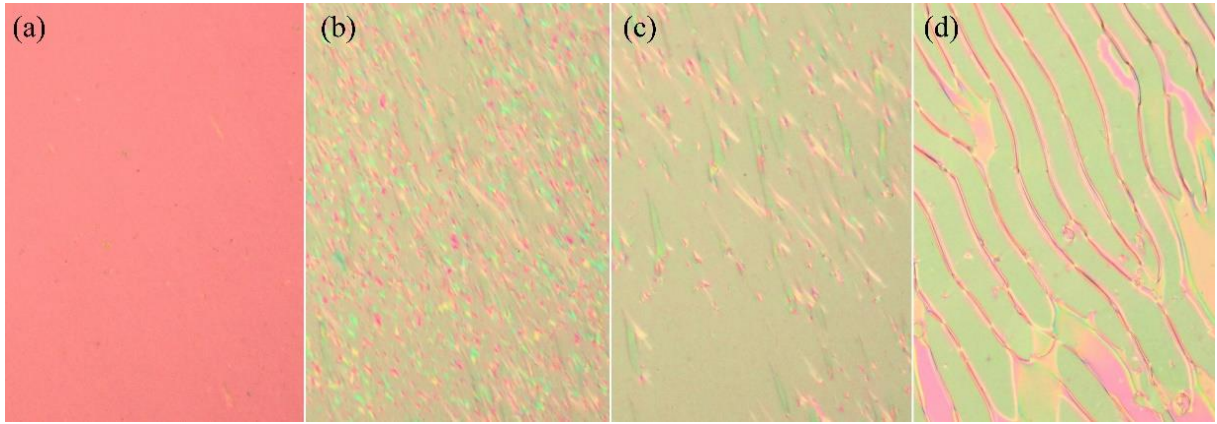

Figure S4. FF6 in a 10-μm HG parallelly rubbed cell in (a) the N phase, (b) at the N-$N_F$ phase transition at T=102.5°C, photos (c) and (d) are taken in the $N_F$ phases at T=95°C and T=80°C, respectively.

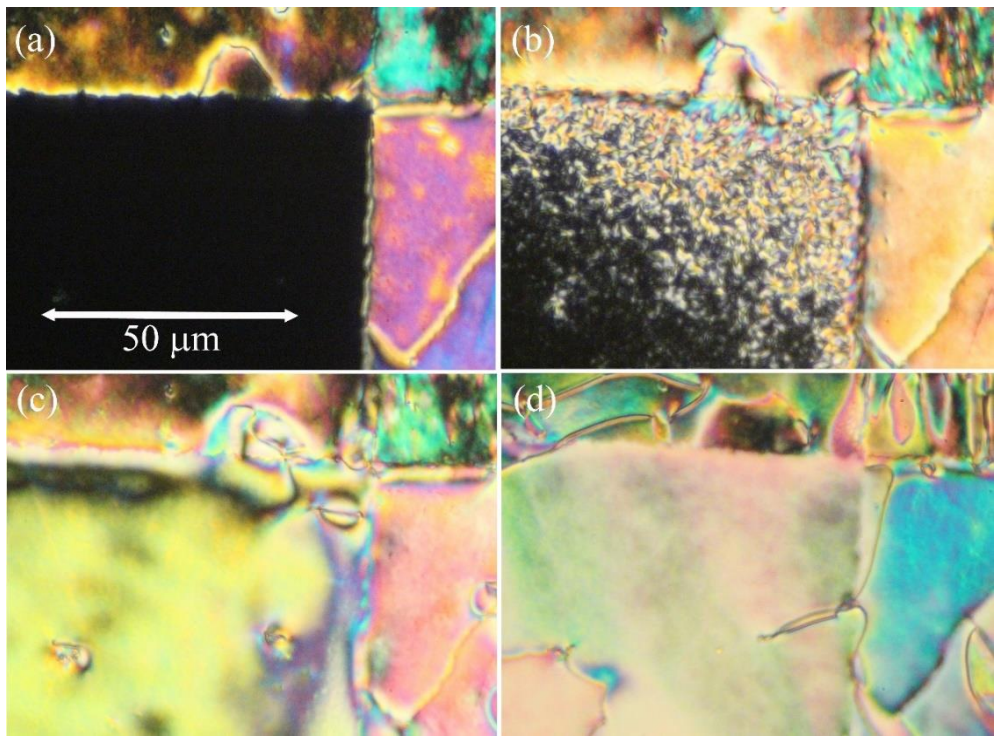

Figure S5. Textures for compound FF6 in a home-made cell without surfactant, (a) in the nematic phase, T=110°C, (b) at the N-$N_F$ phase transition, T=103°C, (c) in the $N_F$ phase, T=102°C and (d) T=100°C. Bottom left part of all figures is the area between electrodes, which is dark in (a) and (b) pictures. Polarizers are oriented along the edges of the photos.

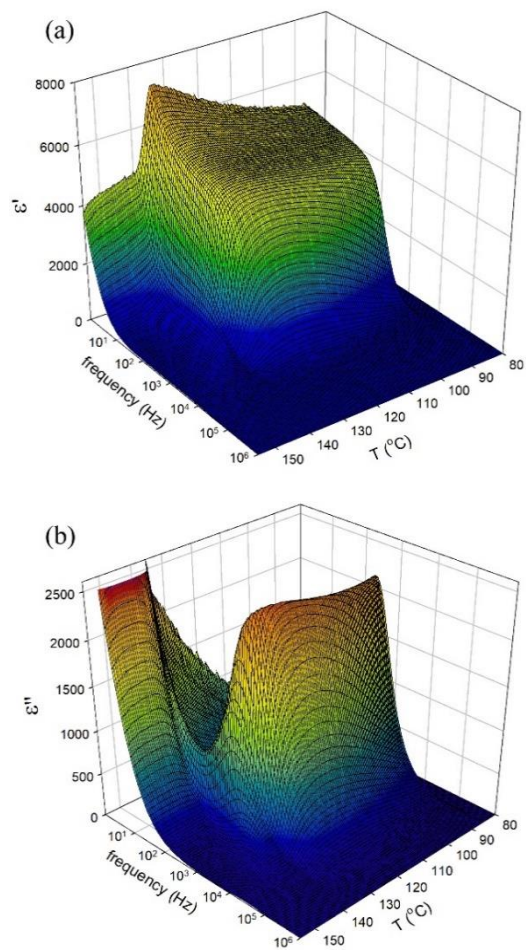

Figure S6. Permittivity versus temperature, $T$, and frequency in 3D-plots: (a) the real, $\varepsilon'$, and (b) the imaginary, $\varepsilon''$, part of permittivity for compound ClF6. Dielectric measurements were performed in a 5-μm cell without surfactant layers.

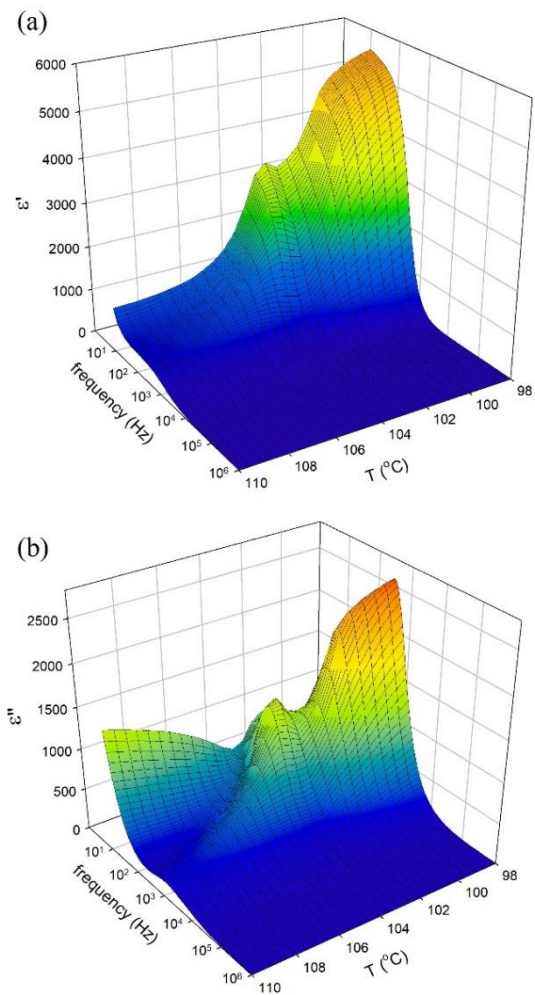

Figure S7.　For FF6 compound, the permittivity versus temperature, $T$, and the frequency dependences in the vicinity of the N-$N_F$ phase transition: (a) the real, $\varepsilon'$, and (b) the imaginary, $\varepsilon''$, part of permittivity. The dielectric measurements were performed in a 5-μm cell without surfactant layers

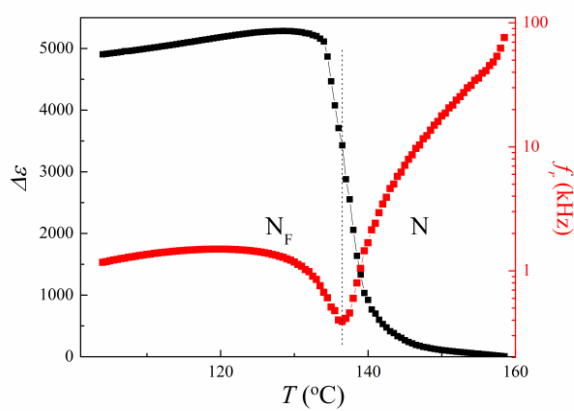

Figure S8.　Temperature dependence of the dielectric strength, $\Delta\varepsilon$, and the relaxation frequency, $f_r$, for compound ClF6. The results of fitting of the dielectric spectroscopy data gained during the measurements in a 5-μm home-made cell without surfactant (Figure S6).

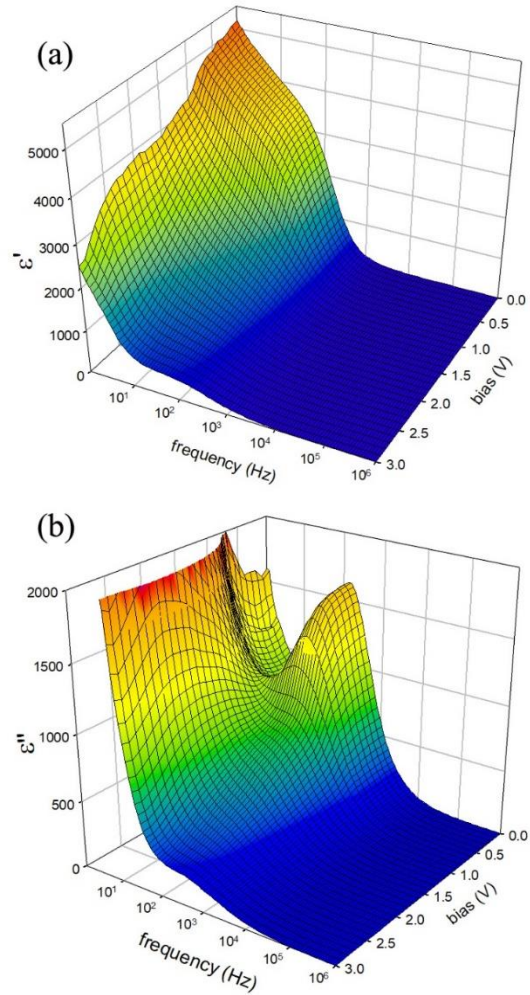

Figure S9. For FF6, the frequency dependences of the (a) real and (b) imaginary parts of the permittivity, $\varepsilon^*$, taken at T=100°C, in the 5-μm cell without surfactant, in bias field from zero to 0.6 V/μm. The measuring field was 0.01 V/μm.

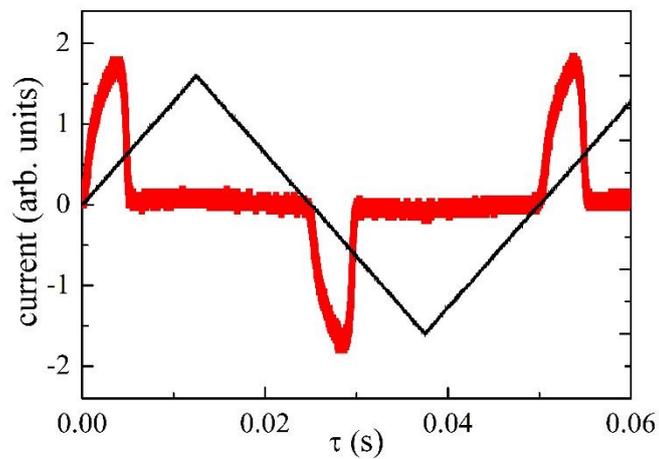

Figure S10. The polarization current (red colour) versus the profile of applied field of about ±1V/mm at 20Hz for FF6 at T=85°C.